\documentclass[11pt]{article}

\usepackage{amsfonts,amssymb,bm,cite,amsmath,braket}

 \usepackage[pdftex]{graphicx}
\usepackage{array, booktabs}
\usepackage[subrefformat=parens]{subcaption}
\usepackage{here}
\newcommand {\beq}{\begin{equation}}
\newcommand {\eeq}{\end{equation}}
\newcommand {\beqa}{\begin{eqnarray}}
\newcommand {\eeqa}{\end{eqnarray}}
\newcommand {\n}{\nonumber \\}
\newcommand {\tr}{\mbox{tr}}
\newcommand {\Tr}{\mbox{Tr}}

\newcommand {\del}{\partial}

\newcommand {\mathH}{\mathcal{H}}

\newcommand {\mathO}{\mathcal{O}}

\renewcommand{\theequation}{\thesection.\arabic{equation}}

\begin{document}

\setlength{\oddsidemargin}{0cm}
\setlength{\baselineskip}{6mm}

\abovedisplayskip=1.0em
\belowdisplayskip=1.0em
\abovedisplayshortskip=0.5em
\belowdisplayshortskip=0.5em

 \titlepage

 \begin{titlepage}
 \renewcommand{\thefootnote}{\fnsymbol{footnote}}
 \begin{normalsize}
 \begin{flushright}
 \begin{tabular}{l}
January 2022
 \end{tabular}
 \end{flushright}
 \end{normalsize}

 ~~\\

 \vspace*{0cm}
     \begin{Large}
        \begin{center}
          {Target space entanglement in a matrix model for the bubbling geometry}
        \end{center}
     \end{Large}
 \vspace{1cm}

 \begin{center}
            Asato T{\sc suchiya}\footnote
             {
 e-mail address :
 tsuchiya.asato@shizuoka.ac.jp}
            and
            Kazushi Y{\sc amashiro}\footnote
            {
 e-mail address : yamashiro.kazushi.17@shizuoka.ac.jp}\\
       \vspace{1.5cm}

 {\it Department of Physics, Shizuoka University}\\
                {\it 836 Ohya, Suruga-ku, Shizuoka 422-8529, Japan}\\
          \vspace{0.5cm}
  {\it Graduate School of Science and Technology, Shizuoka University}\\
                  {\it 836 Ohya, Suruga-ku, Shizuoka 422-8529, Japan}

 \end{center}

 \vspace{3cm}

 \begin{abstract}
 \noindent
We study the target space entanglement entropy in a  complex matrix model 
that describes the chiral primary sector in $\mathcal{N}=4$ super Yang-Mills theory, which is 
associated with the bubbling AdS geometry. The target space for the matrix model
is a two-dimensional plane where the eigenvalues of the complex matrix distribute.
The eigenvalues are viewed as the position coordinates of fermions, and
the eigenvalue distribution corresponds to  a droplet formed by the fermions. The droplet is identified with
one that specifies a boundary condition in the bubbling geometry.
We consider states in the matrix model that correspond to $AdS_5\times S^5$, 
an AdS giant graviton and a giant graviton in the bubbling geometry.
We calculate the target space entanglement entropy of a subregion for each of the 
states in the matrix model as well as
the area of the boundary of the subregion in the bubbling geometry, and find 
a qualitative agreement between them.
 \end{abstract}
 \vfill
 \end{titlepage}


\vfil\eject

\setcounter{footnote}{0}


\section{Introduction}
\setcounter{equation}{0}
It has been recognized that quantum entanglement plays a crucial role in constructing a quantum theory
of gravity. As seen typically in the Ryu-Takayanagi formula\cite{Ryu:2006bv},  quantum entanglement has information on
geometry of space-time. The formula tells that when the space where a field theory is defined is divided into a subregion and its complement, the 
entanglement entropy of the subregion is proportional 
 to the area of minimal surface in the bulk whose
boundary agrees with that of the subregion. 
The entanglement entropy in the formula is here called the base space entanglement entropy. 

Recently, a new type of quantum entanglement entropy in field theories which is 
called the target
space entanglement entropy has been investigated in the context of 
the gauge/gravity correspondence\cite{Mazenc:2019ety,Hampapura:2020hfg,Das:2020jhy,Das:2020xoa,Sugishita:2021vih,Frenkel:2021yql}. 
It is defined by
dividing the configuration space of fields into a subregion and its complement.
In particular, the target space entanglement entropy can be defined in $(0+1)$-dimensional 
field theory, namely quantum mechanics, although the base space entanglement entropy cannot be defined 
there.  For instance, in matrix quantum mechanics, 
the target space is identified with a space where the eigenvalues of the matrix distribute.
It is nontrivial to define entanglement entropy in gravitational theories because of division of dynamical spaces. 
It has been conjectured in \cite{Das:2020jhy} 
that 
in the gauge/gravity correspondence,  the entanglement entropy in the bulk on the gravity side can be defined by the target space entanglement entropy on the gauge theory side.  
In particular, 
the target space entanglement entropy of a subregion
in the D0-brane quantum mechanics\cite{Banks:1996vh}
is expected to be proportional with factor of proportionality $1/4G_N$ 
to the area of boundary of the corresponding subregion in the bulk on the gravity side.

In this paper, we study this issue in the D3-brane holography, namely a conjectured correspondence between
$\mathcal{N}=4$ super Yang-Mills theory (SYM) and type IIB superstring theory on $AdS_5\times S^5$\cite{Maldacena:1997re,Gubser:1998bc,Witten:1998qj}.
Here we focus on a correspondence between 
the chiral primary sector and the bubbling AdS geometry\cite{Corley:2001zk,Berenstein:2004kk,Lin:2004nb}.
The chiral primary sector can be described by a complex matrix model\cite{Takayama:2005yq} whose 
target space is a two-dimensional plane where the eigenvalues of the complex matrix distribute (see also \cite{Ghodsi:2005ks}).
The eigenvalues can be viewed as the position coordinates of fermions in the harmonic oscillator 
potential in the two-dimensional plane, and the eigenvalue distribution corresponds to a droplet formed
by the fermions.
The target space can be identified with a two-dimensional plane in the bubbling geometry
where a boundary condition for a function that determines a half-BPS solution with
$R \times SO(4) \times SO(4)$ symmetry in type IIB supergravity
is specified by giving a droplet. 
We calculate the target space entanglement entropy of a subregion in the two-dimensional plane
for each of states that are specified by droplets corresponding to $AdS\times S^5$,  an AdS giant graviton and 
a giant graviton as well as the area (length) 
of boundary of the corresponding subregion
in the bubbling geometry. We find a qualitative agreement between the target space entanglement
entropy and the area.

This paper is organized as follows.
In section 2, we review some materials that we will need in this paper:
a complex matrix model that describes the chiral primary sector in $\mathcal{N}=4$ SYM on $R\times S^3$,
the bubbling AdS geometry, and the target space entanglement entropy.
In section 3, we calculate the entanglement entropy in the complex matrix model.
In section 4, we calculate the area of boundary in the bubbling geometry
and compare it with the target space entanglement entropy.
Section 5 is devoted to conclusion and discussion.
In appendices, some details are gathered.

\section{Review}
\setcounter{equation}{0}

\subsection{Complex matrix model and fermions in two-dimensional plane }
In this subsection, we review a complex matrix model that describes the chiral primary sector of $\mathcal{N}=4$ SYM on $R\times S^3$.
The chiral primary operators take the form
\begin{align}
  \mathO^{J_1, J_2,\cdots , J_K}(t) = \prod^K_{a=1} \Tr (Z^{J_a}) \ ,
  \label{half BPS op}
\end{align}
where $Z = \frac{1}{\sqrt{2}} (\phi_1 + i\phi_2)$ with $\phi_1$ and $\phi_2$
being two of six scalars in $\mathcal{N}=4$ SYM.
These operators are half-BPS holomorphic operators.
Kaluza-Klein gravitons, AdS giant gravitons and giant gravitons on the gravity side
are represented in terms of a linear combination of the operators (\ref{half BPS op}).
It was shown in \cite{Takayama:2005yq} that the dynamics of the operators is described 
by a complex matrix model which is obtained
by dimensionally reducing the free part of the action for $Z$ on $R\times S^3$ to $R$.
The complex matrix model is a matrix quantum mechanics defined by
\begin{align}
  \mathcal{Z} &= \int [dZ(t)dZ^{\dagger}(t)]e^{iS}   \ , \n
  S &= \int dt \Tr (\dot{Z}(t) \dot{Z^{\dagger}}(t) - Z(t)Z^{\dagger}(t)) = \int dt\sum_{i,j}  (\dot{Z}(t)_{ij} \dot{Z^{*}}(t)_{ij} - Z(t)_{ij}Z^{*}(t)_{ij}) \ , 
  \label{CM}
\end{align}
where $Z(t)$ is an $N \times N$ complex matrix depending on the time, and  the 
path integral measure is defined by a norm in matrix configuration space,
\begin{align}
  ||dZ(t)||^2 = 2\Tr(dZ(t)dZ^{\dagger}(t)) = 2 \sum_{i,j} dZ(t)_{ij} dZ^*(t)_{ij}  \ .
\end{align}
The potential term in the action comes from the coupling of the conformal matter to the curvature
of $S^3$. We have rescaled the field and the time appropriately.

The hamiltonian is given by
\begin{align}
  \hat{H} = \sum_{i,j} \left( -\frac{\del}{\del Z_{ij} \del Z^*_{ij}} + Z_{ij} Z^*_{ij} \right) \ .
\end{align}
The normalized wave function of the ground state is given by
\begin{align}
  \chi_0 = \frac{1}{\pi^{\frac{N^2}{2}}} e^{-\Tr(ZZ^{\dagger})} = \frac{1}{\pi^{\frac{N^2}{2}}} e^{\Sigma_{i,j} Z_{ij} Z^*_{ij}}  \ ,
\end{align}
and the wave functions of excited states that correspond to the chiral primary states are given by
\begin{align}
  \chi^{(J_1, \cdots, J_K)} = \left( \prod^{K}_{a=1} \Tr(Z^{J_a}) \right) \chi_0  \ .
  \label{CM wf}
\end{align}
The energy eigenvalues of these states are $N^2 + \Sigma^K_{a=1}J_a$.

The complex matrix $Z$ is decomposed  as $Z = U T U^{\dagger}$
in terms of a unitary matrix $U$ and an upper  triangular matrix $T$.
The eigenvalues of $Z$ are given by $z_i = T_{ii} \; (i=1,\cdots,N)$. 
The wave functions (\ref{CM wf}) are rewritten in terms of $z_i$ and $T_{ij} \; (i<j)$ as
\begin{align}
  \chi^{(J_1, \cdots, J_K)} &= \left( \prod^{K}_{a=1} \sum_{i_a} z^{J_a}_{i_a} \right) \chi_0 \ ,  \n
  \chi_0 &= \frac{1}{\pi^{\frac{N^2}{2}}} e^{-\Sigma_i z_{i} z^*_{i} - \Sigma_{j<k} T_{jk} T^{*}_{jk} } \ ,
\end{align}
while the path integral measure as 
\begin{align}
  \int \prod_{i,j} dZ_{ij} dZ^*_{ij}  = \int \prod_{i>j} dH_{ij} dH^*_{ij} \prod_{k<l} dT_{kl} dT^*_{kl} \prod_m dz_m dz^*_m |\Delta(z)|^2 \ ,
\end{align}
where $\Delta(z) = \prod_{i<j} (z_i -z_j)$ and $dH = -i U^{\dagger} dU$.
We can absorb $\Delta(z)$ into the wave function and 
define a new wave function $\chi_F$ by $\chi_F \equiv \Delta(z) \chi$.
Then, $\chi_F$ is represented as a certain linear combination of
\begin{align}
&  \frac{1}{\sqrt{N!}}\det (\Phi_{l_i}(z_j, z^*_j)) \times \prod_{j<k} \Phi_0(T_{jk}, T^*_{jk}) \  ,  \n
& \mbox{with} \;\; \sum_i l_i = \frac{N(N-1)}{2} + \sum_{a=1}^K J_a  \ ,
  \label{CM wf 2}
\end{align}
where  $\Phi_{l}(z, z^*)$ are the wave function
of the lowest Landau level which takes the form
\begin{align}
  \Phi_{l}(z, z^*)= \sqrt{\frac{2^l}{l! \pi}} z^l e^{-z z^*}  \ .
  \label{LLL}
\end{align}
As reviewed in appendix A, 
the wave function $\Phi(z,z^*)$ can be viewed as a wave function of the holomorphic sector
for a particle in the harmonic oscillator potential in two-dimensional plane.
(\ref{CM wf 2}) is a wave function for $N$ fermions and $\frac{1}{2}N(N-1)$ bosons
in the harmonic oscillator potential.
The eigenvalues $z_i$ of $Z$ can  be viewed as the coordinates of
$N$ fermions and  the total energy of these fermions is 
$\frac{1}{2}N(N+1) + \Sigma^K_{a=1}J_a$, where $\frac{1}{2}N(N+1)$ is the energy of the ground state
of the fermions.
On the other hand, the bosons whose coordinates are $(T_{ij},T_{ij}^*)$ are always in the ground state.
Hence, we can integrate out $T_{ij}$ and concentrate on $\frac{1}{\sqrt{N!}}\det (\Phi_{l_i}(z_j, z^*_j))$
in (\ref{CM wf 2}).
In this way, the holomorphic sector of the complex matrix model, 
which describes the dynamics
of the operators (\ref{half BPS op}), reduces to a system of $N$ fermions 
in the two-dimensional harmonic oscillator potential. 
For large $N$, we can view the eigenvalue distribution corresponding to a state as a droplet in the complex plane, which is formed by $N$ fermions.
\begin{figure}[t]
  \begin{tabular}{ccc}
  \begin{minipage}[b]{0.3\linewidth}
    \centering
    \includegraphics[keepaspectratio, scale=0.7]{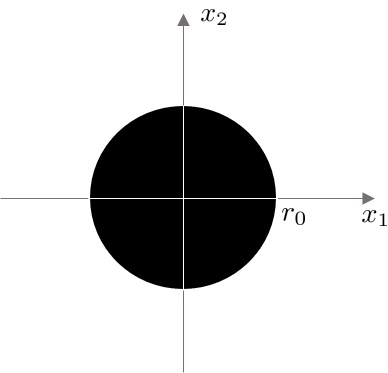}
    \subcaption{$AdS_5 \times S^5$}
    \label{Droplet_AdS}
  \end{minipage}&
  \begin{minipage}[b]{0.3\linewidth}
    \centering
    \includegraphics[keepaspectratio, scale=0.7]{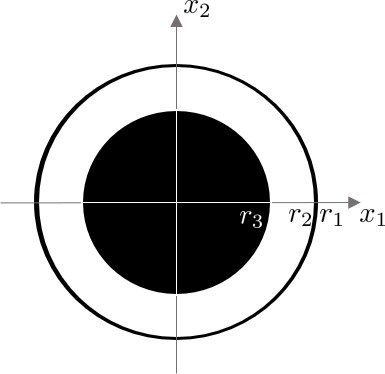}
    \subcaption{AdS giant graviton}
    \label{Droplet_AdS_giant}
  \end{minipage}&
  \begin{minipage}[b]{0.3\linewidth}
    \centering
    \includegraphics[keepaspectratio, scale=0.7]{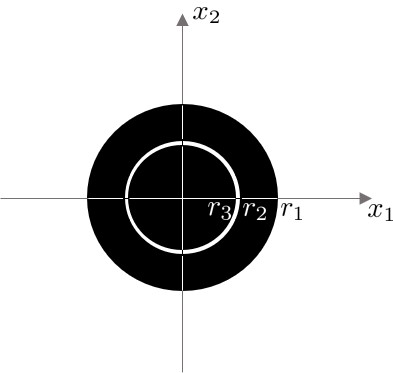}
    \subcaption{giant graviton}
    \label{Droplet_giant}
  \end{minipage}\\
  \addlinespace[3mm]
  \begin{minipage}[b]{0.3\linewidth}
    \centering
    \includegraphics[keepaspectratio, scale=0.5]{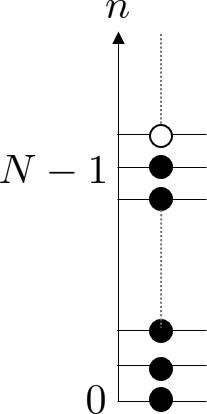}
    \subcaption{$AdS_5 \times S^5$}
    \label{level_AdS}
  \end{minipage}&
  \begin{minipage}[b]{0.3\linewidth}
    \centering
    \includegraphics[keepaspectratio, scale=0.5]{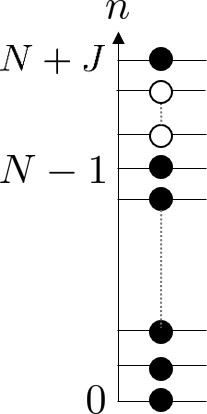}
    \subcaption{AdS giant graviton}
    \label{level_AdS_giant}
  \end{minipage}&
  \begin{minipage}[b]{0.3\linewidth}
    \centering
    \includegraphics[keepaspectratio, scale=0.5]{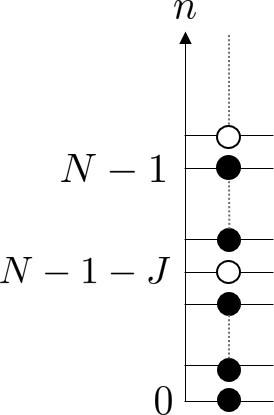}
    \subcaption{giant graviton}
    \label{level_giant}
  \end{minipage}
  \end{tabular}
  \caption{(\ref{Droplet_AdS}), (\ref{Droplet_AdS_giant}) and (\ref{Droplet_giant}) are droplets of
  the states that 
  correspond to $AdS_5\times S^5$, an AdS giant graviton and a giant graviton, respectively. 
  (\ref{level_AdS}),  (\ref{level_AdS_giant}) and  (\ref{level_giant}) are the occupied energy levels
  of the states that correspond to $AdS_5\times S^5$, an AdS giant graviton and a giant graviton, respectively. }
  \label{droplets}
 \end{figure}

In what follows, we consider three particular states whose wave functions are given by
\begin{align}
\frac{1}{\sqrt{N!}}\det (\Phi_{l_i}(z_j, z^*_j))
\end{align}
in (\ref{CM wf 2}).
The first one is the ground state where $l_1=0,\ l_2=1,\ \cdots,\ l_N=N-1$.
We show the corresponding droplet in Fig. \ref{Droplet_AdS} and the occupied energy levels in 
Fig. \ref{level_AdS}.
This corresponds to $AdS_5\times S^5$ in the bubbling geometry.
The second one is an excited state where $l_1=0,\ l_2=1,\ \cdots,\ l_{N}= N-1,\ l_{N+1}=N+J$.
Note that the total number of fermions is $N+1$.
We show the corresponding droplet in Fig. \ref{Droplet_AdS_giant} and the occupied energy levels
in Fig. \ref{level_AdS_giant}.
This corresponds to a bubbling geometry where there exists  
an AdS giant graviton, which is a D3-brane wrapped on $S^3$ in $AdS_5$.
The third one is an excited state where $l_1=0,\ l_2=1,\ \cdots,\ l_{N-1-J}= N-2-J,\ l_{N-J}=N-J,\ \cdots,\ l_{N-1}=N-1$. Note that the total number of fermions is $N-1$.
We show the corresponding droplet in Fig .\ref{Droplet_giant} and the occupied energy levels
in Fig. \ref{level_giant}.
This corresponds to a bubbling geometry where there exists a giant graviton, 
which is a D3-brane wrapped on $S^3$ in $S^5$.
The wave function $\Phi_l(z,z^*)$ is localized on a circle centered at the origin with the radius $\sqrt{l}$.
Hence, $r_0$ in Fig. \ref{Droplet_AdS} is equal to $\sqrt{N}$ for large $N$.

\subsection{Bubbling geometry}
In this subsection, we review the bubbling geometry  developed by
Lin, Lunin and Maldacena. They gave the general form of  half-BPS solutions with $R\times SO(4)\times SO(4)$ 
symmetry in type IIB supergravity \cite{Lin:2004nb}.
The metric takes the form
\begin{align}
  ds^2 = -h^{-2}\left[ dt + \sum_{i=1}^{2} V_i dx^i\right]^2 + h^2 \left[ dy^2 + \sum_{i=1}^2 dx^i dx^i \right] + y e^G d\Omega_3^2 + y e^{-G} d \tilde{\Omega}^2_3  \ , 
  \label{LLM metric}
\end{align}
where 
\begin{align}
  h^{-2} = 2y \cosh G , \ \ 
z = \frac{1}{2}\tanh G, \ \ 
  y \del_y V_i = \epsilon_{ij}\del_j z, \ \ 
y(\del_i V_j - \del_j V_i) = \epsilon_{ij} \del_y z    \ .
\label{func h V}
\end{align}
The function $z(x_1,x_2,y)$ completely determines the solutions and obeys
a differential equation
\begin{align}
  \del_i \del_i z + y \del_y \left(\frac{\del_y z }{y}\right) = 0 \ .
  \label{differential equation}
\end{align}
The function $z$ is fixed by specifying a boundary condition at $y=0$.
It turns out that $z$ must take $1/2$ or $-1/2$ at $y=0$.
It is convenient to assign `black' to the region with $z=-1/2$ and `white' to the region with
$z=1/2$. 
In this way, there is a correspondence between divisions of the $x_1-x_2$ plane
 into black and white regions and half-BPS solutions with $R\times SO(4) \times SO(4)$ symmetry.
One can identify a two-dimensional plane
specified by $y=0$ in the bubbling geometry with the target space in the complex matrix model
such that black regions correspond to droplets in the target space.
Figs \ref{Droplet_AdS}, \ref{Droplet_AdS_giant} and \ref{Droplet_giant} correspond to 
$AdS_5\times S^5$, an AdS giant graviton and a giant graviton, respectively, as mentioned before.
Here $r_0$ in Fig.\ref{Droplet_AdS} is related to the radius of $AdS_5$ as $r_0 = R_{AdS}^2= \sqrt{N}$.

\subsection{Target Space Entanglement Entropy}
In this subsection, we briefly review the target space entanglement entropy\cite{Mazenc:2019ety}.
For concreteness, we consider a quantum mechanics of $N$ identical particles in $d$ dimensions.
In this case, the target space is the $d$-dimensional space where $N$ particle reside.
We divide the target space into a subregion $A$ and its complement $\bar{A}$.
Then, the total Hilbert space $\mathcal{H}$ is decomposed as
\begin{align}
  \mathH = \bigoplus_n \mathH_{A,n} \otimes \mathH_{\bar{A},n}  \ ,
  \label{direct sum}
\end{align}
where $n$ labels a sector where $n$ particles exist in $A$ and $N-n$ 
particles exist in $\bar{A}$.
The target space entanglement entropy of the subregion $A$ is given by
\begin{align}
  S_A = - \sum_n p_n \log p_n + \sum_n p_n S_{A,n} \ .
  \label{TSEE}
\end{align}
The first and second terms in the RHS is  classical and quantum parts, respectively.
Here $p_n$ is the probability of realization of the sector $n$, which is given by
\begin{align}
  p_n &= \begin{pmatrix}
    N \\
    n \\
    \end{pmatrix}   
    \int_A \prod_{a=1}^nd^dx_a \int_{\bar{A}} \prod_{b=n+1}^{N} d^dy_b
     | \psi (\vec{x}_1, \cdots, \vec{x}_n, \vec{y}_{n+1}, \cdots , \vec{y}_N)|^2 \ .
\label{p_i}
\end{align}
$S_{A,n}$ is interpreted as the entanglement entropy of the subregion $A$ in the sector $n$,
which is given by
\begin{align}
  S_{A,n} = -\tr_{A,n} \rho_{A,n} \log \rho_{A,n} \ ,
  \label{S_Ai}
\end{align}
where $\rho_{A,n}$ is the reduced density matrix of the sector $n$ and its 
matrix elements in the coordinate basis are given by
\begin{align}
  \bra{\vec{x}} \rho_{A,n} \ket{\vec{x}'} =
  \frac{1}{p_n}
  \begin{pmatrix}
    N \\
    n \\
    \end{pmatrix}
    \int_{\bar{A}} \prod_{b=n+1}^Nd^{d}y_b \psi(\vec{x}, \vec{y}) \psi^*(\vec{x'}, \vec{y})  \ .
\label{rho_Ai}
\end{align}

As a particular example\cite{Sugishita:2021vih}, we consider $N$ identical fermions which are not interacting each other and 
the wave function of which is given by a Slater determinant
\begin{align}
  \psi(\vec{x}_1 , \cdots \vec{x}_N) 
  = \frac{1}{\sqrt{N!}} \det (\chi_i(\vec{x}_j))  \ ,
\end{align}
where $\chi_i(\vec{x})$ are orthonormal single-body wave functions:
\begin{align}
\int d^d x \chi_i^*(\vec{x}) \chi_j(\vec{x}) = \delta_{ij} \ .
\end{align}
We introduce an $N\times N$ overlap matrix $X_{ij}$
\begin{align}
  X_{ij} \equiv \int_A d^d x  \  \chi_i(\vec{x})\chi^*_j(\vec{x})    \ , 
 \label{overlap mat}
\end{align}
$X_{ij}$ is a hermitian matrix so that it is diagonalized by a unitary matrix $U_{ij}$.
The eigenvalues of $X_{ij}$ denoted by $\lambda_i$ are given by
\begin{align}
  \lambda_i = \int_A | \tilde{\chi}_i (\vec{x})^2 |   \ ,
\end{align}
where  $\tilde{\chi}_i(\vec{x}) = U_{ij} \chi_j (\vec{x})$.
$\lambda_i$ is the probability of existence in the region $A$ when the wave function for a particle is given by $\tilde{\chi}_i(\vec{x})$.

 $p_n$ in (\ref{p_i}) is given by
\begin{align}
  p_n = \sum_{I\in F_n} \prod_{i\in I} \lambda_i \prod_{j \in \bar{I}} (1 - \lambda_j) \ .
\end{align}
Here $F_n$ is a set of all subsets of $\{1,\ldots,N\}$ that consist of $n$ elements\footnote{For instance , in the case of
$N=3, n = 2$,  $F_2 = \{\{1,2\}, \{1,3\}, \{2,3\}\}$.}. We see that
$p_n$ is indeed the probability that $n$ particles
exist in the region $A$ when the probability that each of $N$ particles exists in $A$ is given by $\lambda_i$.

The matrix elements of $\rho_{A,n}$ in (\ref{rho_Ai}) are 
\begin{align}
  \bra{\vec{x}} \rho_{A,n} \ket{\vec{x}'} 
    = \frac{1}{p_n} \sum_{I\in F_n} \lambda_I \bar{\lambda}_I \psi_I(\vec{x}) \psi_I^*(\vec{x'}) \ ,
    \label{TSEE cl result}
\end{align}
where we put $\lambda_I \equiv \prod_{i\in I} \lambda_i, \;\;\bar{\lambda}_I \equiv \prod_{i\in I} (1-\lambda_i)$ for an element $I$ of $F_n$ and introduce a wave function for $n$ particles
\begin{align}
  \psi_I = \frac{1}{\sqrt{\lambda_I}} \sum_{\sigma \in S_n} \frac{(-1)^{\sigma}}{\sqrt{n!}}\tilde{\chi}_{n_{\sigma(1)}}(\vec{x}_1) \cdots \tilde{\chi}_{n_{\sigma(n)}}(\vec{x}_n)  \ . 
\end{align}
The quantum part is calculated as
\begin{align}
  \sum^N_{n=0} p_n S_{A,n} = \sum^N_{n=0} p_n \log p_n - \sum^N_{n=0} \sum_{I \in F_n} \lambda_I \bar{\lambda}_{I} \log (\lambda_I \bar{\lambda}_{I})  \ .
  \label{TSEE q result}
\end{align}

Thus, we see from (\ref{TSEE cl result}) and (\ref{TSEE q result}) that
 the total target space entanglement entropy (\ref{TSEE}) is given by
\begin{align}
  S_A = - \sum^N_{n=0} \sum_{I \in F_n} \lambda_I \bar{\lambda}_{I} \log (\lambda_I \bar{\lambda}_{I}) .
  \label{TSEE result}
\end{align}
By introducing $H(\lambda) \equiv - \lambda \log \lambda - (1-\lambda) \log (1-\lambda)$,
we represent (\ref{TSEE result}) as\footnote{This formula was obtained in
\cite{Klich:2004pb,Rodriguez:2008yql,Calabrese:2011zzb} in the context of condensed matter and
statistical physics.}
\begin{align}
  S_A = \sum_{i=1}^N H(\lambda_i) \ .
  \label{H}
\end{align}
Here $H(\lambda)$ is the Shannon entropy for the Bernoulli distribution $(\lambda, 1-\lambda)$.

\section{Target Space Entanglement Entropy in the complex matrix model}
\setcounter{equation}{0}
In this section, we calculate the target space entanglement entropy for the three states 
in the complex matrix model introduced in section 2.1.
We consider a circle centered at the origin with the radius $r$ as a subregion $A$.
Then, the overlap matrix (\ref{overlap mat}) is given by
\begin{align}
  X_{l l'} (A ) &= 2 \int_A dz dz^*  \Phi_l(z, z^*) \Phi_{l'}(z, z^*) \n
  & = \delta_{ll'} a_l(r)  \ .
\end{align}
Note that this matrix is diagonal and its eigenvalues $\lambda_l$ are
\begin{align}
  \lambda_l = a_l(r) = \frac{\gamma[l+1, r^2]}{\Gamma[l + 1]}  \ .
\end{align}
$\lambda_l$ is the provability of existence in $A$ of a single particle.
Here $\Gamma[x]$ is the gamma function and $\gamma[a, x]$ is the incomplete gamma function:
\begin{align}
  \gamma[a, x] = \int_0^x t^{a-1} e^{-t} dt  \ .
\end{align}
By using (\ref{TSEE result})\footnote{Here we simply consider the full Hilbert space for $N$ fermions in two-dimensional plane in calculating the reduced density matrix.}, we obtain the target space entanglement entropy of the subregion $A$:
\begin{align}
  S(r,N) = \sum^N_{i=1} H(a_{l_{(i)}}(r)), \ \ \  H(a_{l_{(i)}}(r)) = -a_{l_{(i)}}(r) \log a_{l_{(i)}}(r) -(1-a_{l_{(i)}}(r)) \log (1- a_{l_{(i)}}(r))  \ .
  \label{TSEE for general Slater determinant}
\end{align}

\begin{figure}[h]
  \centering
  \includegraphics[width=11cm]{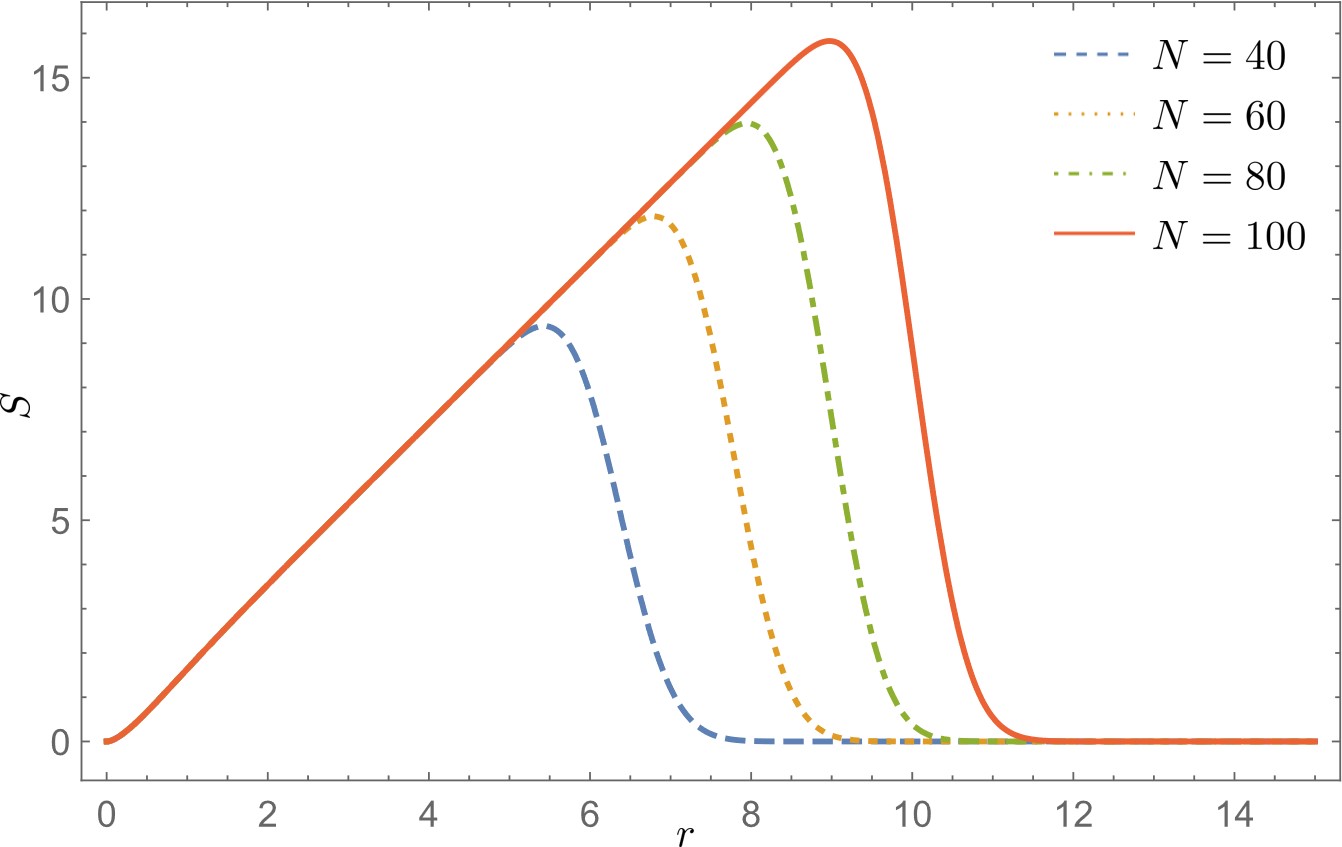}
  \caption{The target space entanglement entropy for the ground state $S_0$ with $N=40, \  60, \  80, \ 100$ are plotted against the radius $r$ of the subregion $A$.}
  \label{fig:TSEE_ground_state} 
\end{figure}

First, let us consider the target space entanglement entropy
for the ground state indicated in Figs. \ref{Droplet_AdS} and \ref{level_AdS}, 
which correspond to $AdS_5\times S^5$ in the bubbling geometry.
In this case, the single-body states with $l=0, \cdots, N-1$ are occupied.
Thus, (\ref{TSEE for general Slater determinant}) is
\begin{align}
  S_0(r, N) = \sum^{N-1}_{l=0} H(a_l(r)), \ H(a_l(r)) = -a_l(r) \log a_l(r) -(1-a_l(r)) \log (1- a_l(r)) \ .
\end{align}
In Fig. \ref{fig:TSEE_ground_state},   we plot the target space entanglement entropy $S_0$ for
$N=40, \ 60,   \ 80, \ 100$ against $r$.
We see that $S_0$ is proportional to $r$ as
$S_0 = 1.81 r$ when the subregion $A$ is included in the droplet,
namely $r < \sqrt{N}$.
This behavior shows that the entanglement entropy is proportional to the area of boundary of  
the subregion $A$.
The fact that $S_0=0$ 
when the droplet is included in the subregion $A$, namely $r>\sqrt{N}$, shows that
there is no entanglement between inside and outside of $A$ because all particles are
confined in $A$. 
This is a finite $N$ effect\footnote{In \cite{Rodriguez:2008yql}, $S_0$ was calculated in the $N\rightarrow\infty$ limit.
The result is that
$S_0 = 2\sqrt{2} \pi r \int^{\infty}_{-\infty} \frac{d\mu}{2\pi} H(\frac{1}{2} {\rm{Erfc}}(\mu)) \sim 1.804 r$ for $0\leq r < \infty$.}
.

\begin{figure}[t]
  \centering
  \includegraphics[width=11cm, page=11]{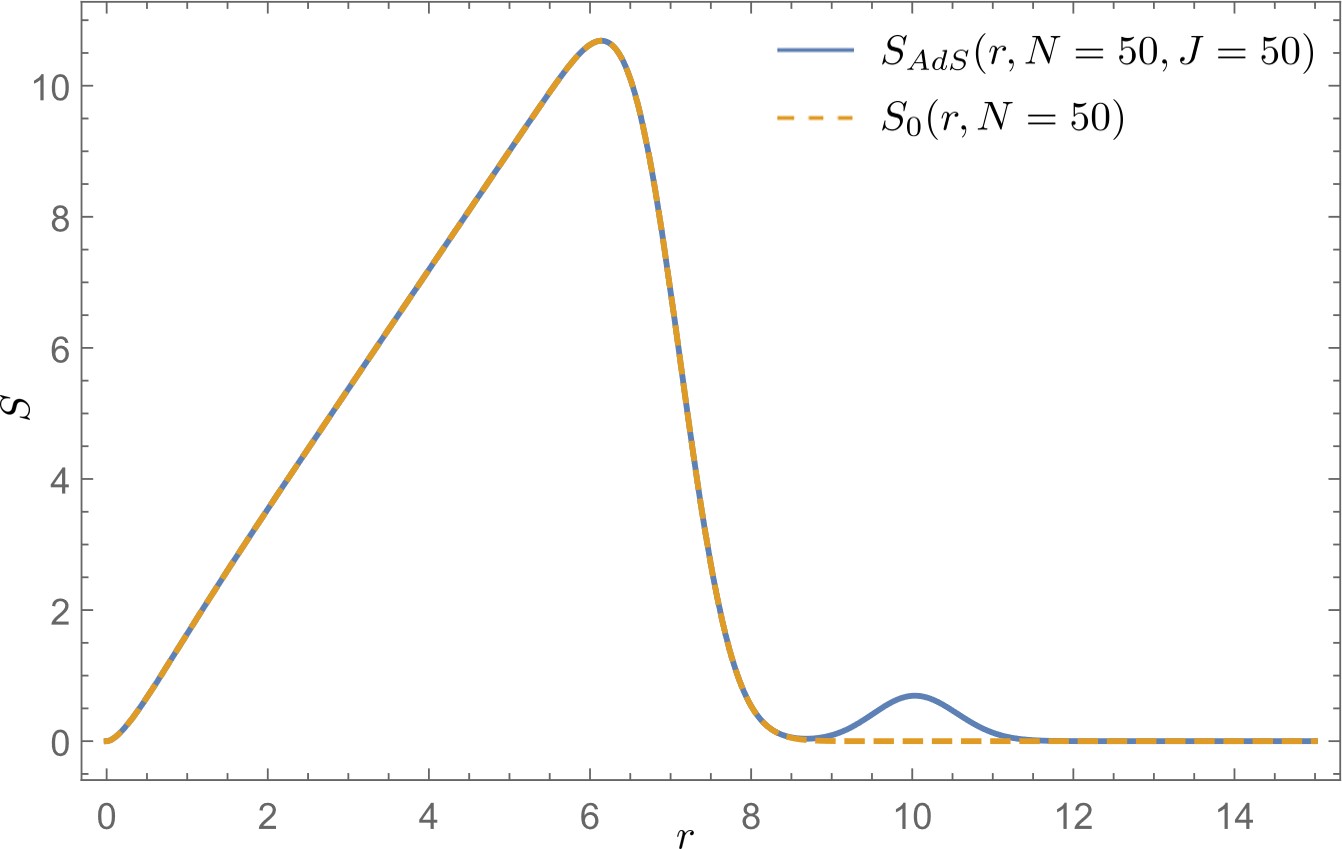}
  \caption{The solid line represents the target space entanglement entropy for the state corresponding to the
  AdS giant graviton $S_{AdS}(r)$ with $N=50$ and $ J=50$, while the dashed line represents the target space
  entanglement entropy for the ground state $S_{0}(r)$ with $N=50$.}
  \label{fig:TSEE_ads_giant}
 \end{figure}

\begin{figure}[h]
   \centering
   \includegraphics[width=11cm,page=11]{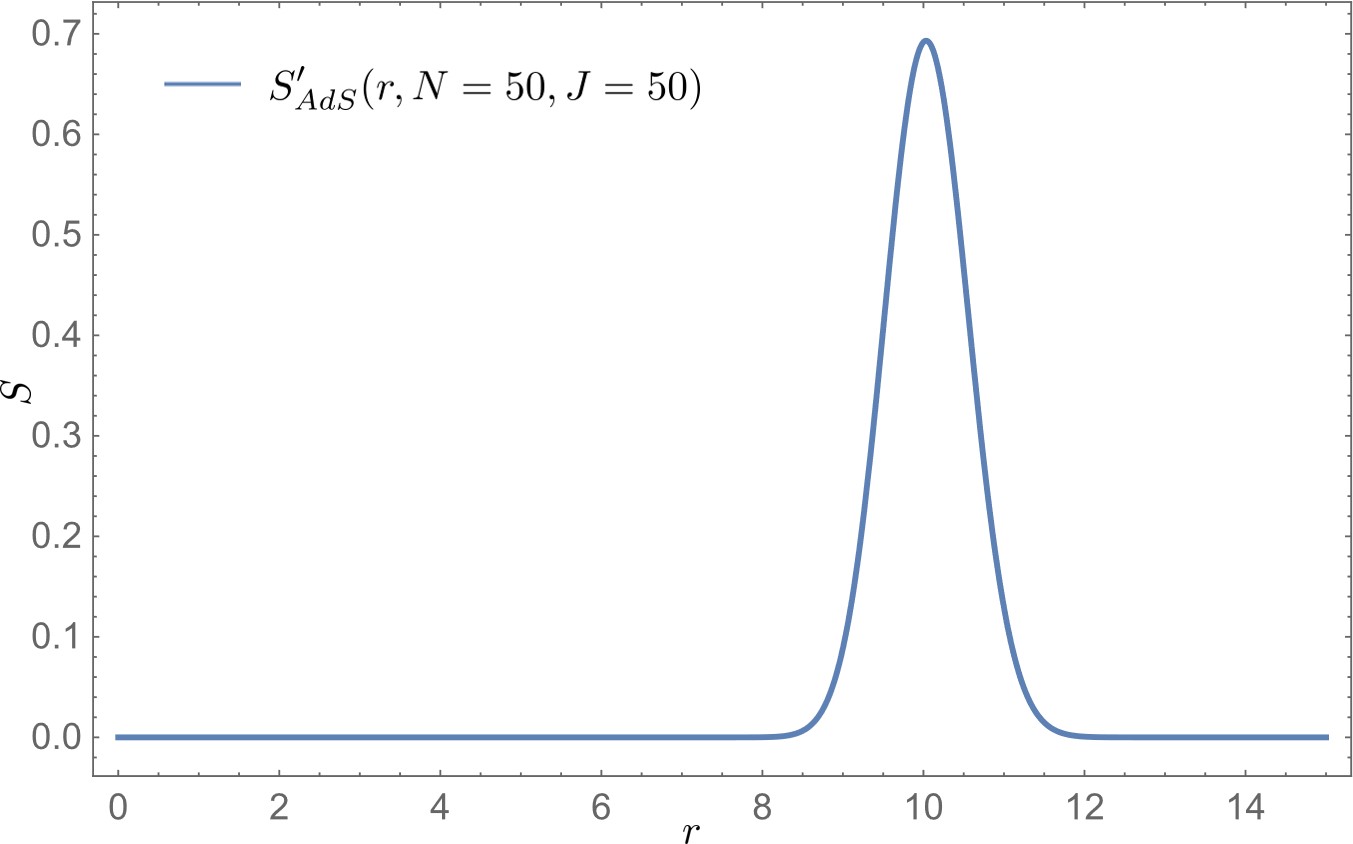}
   \caption{
    $S_{AdS}' = S_{AdS}(r,50,50)-S_0(r,50)$ is plotted against $r$.}
   \label{fig:TSEE_d_ads_giant}
  \end{figure}

Second, we calculate the target space entanglement entropy for an excited state corresponding
to an AdS giant graviton indicated in Figs. \ref{Droplet_AdS_giant} and \ref{level_AdS_giant}.
(\ref{TSEE for general Slater determinant}) in this case is
\begin{align}
  S_{AdS}(r,N,J) = \sum^{N-1}_{i=0} H(a_i(r)) + H(a_{N+J}(r)) \ .
\end{align}
In Fig. \ref{fig:TSEE_ads_giant}, we plot $S_{AdS}$ for $N=50$ and $J=50$ against $r$.
In order to see only the contribution of the AdS giant graviton to the target space entanglement entropy, we subtract $S_0$ from $S_{AdS}$.
In Fig. \ref{fig:TSEE_d_ads_giant}, we plot 
 $S_{AdS}' = S_{AdS}(r,50,50)-S_0(r,50)$ against $r$.
Note here that $N$ for $S_{AdS}$ is different from that for $S_0$. 
We see that there is a peak at $r=\sqrt{N+1+J}$, which is considered as the position of
the AdS giant graviton.

\begin{figure}[t]
  \centering
  \includegraphics[width=11cm]{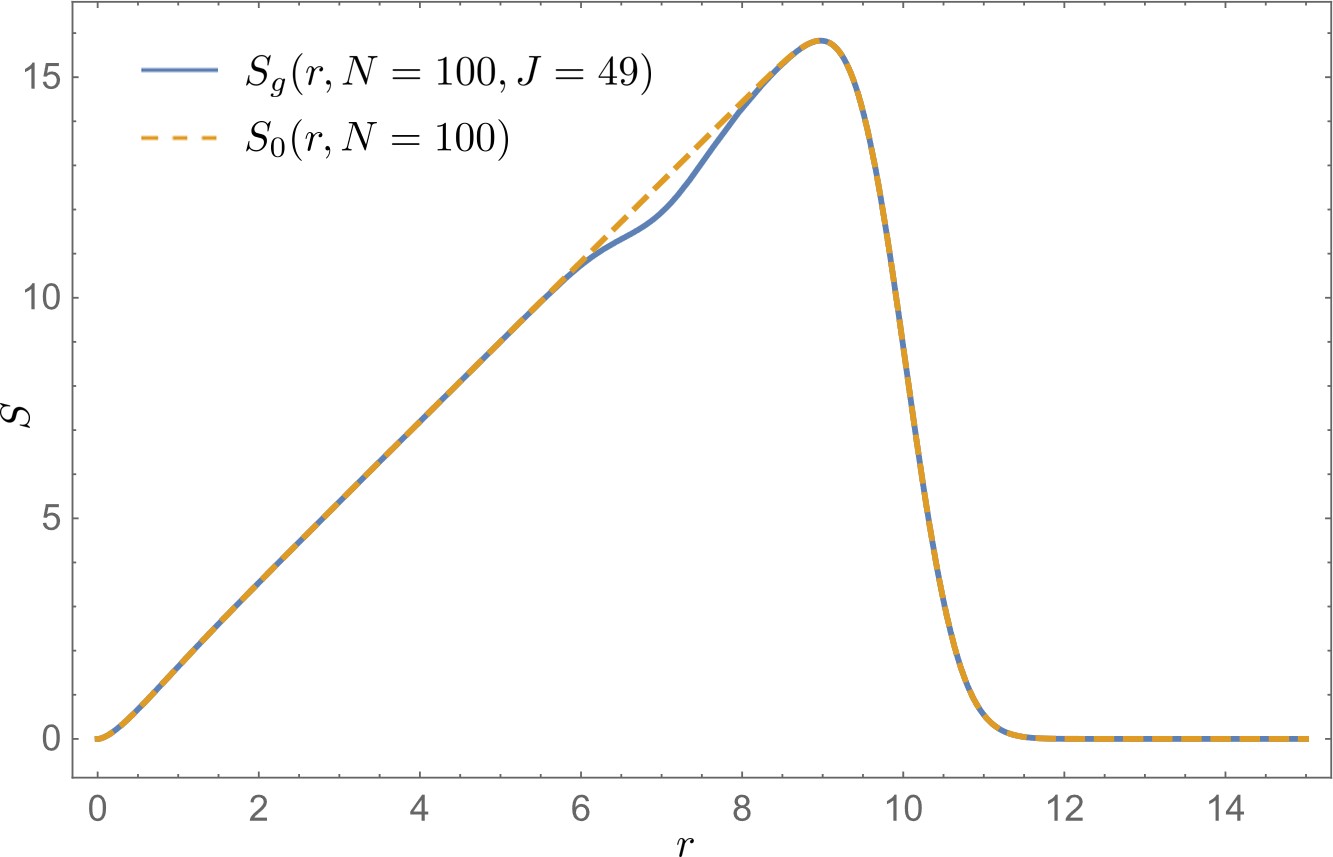}
  \caption{The solid line represents the target space entanglement entropy for the state corresponding
  to the giant graviton $S_{g}(r)$ with $N=100$ and $J=49$, while the dashed line represents the target
  space entanglement entropy for the ground state $S_{0}$ with $N=100$.}
  \label{fig:TSEE_giant}
 \end{figure}

\begin{figure}[h]
  \centering
  \includegraphics[width=11cm]{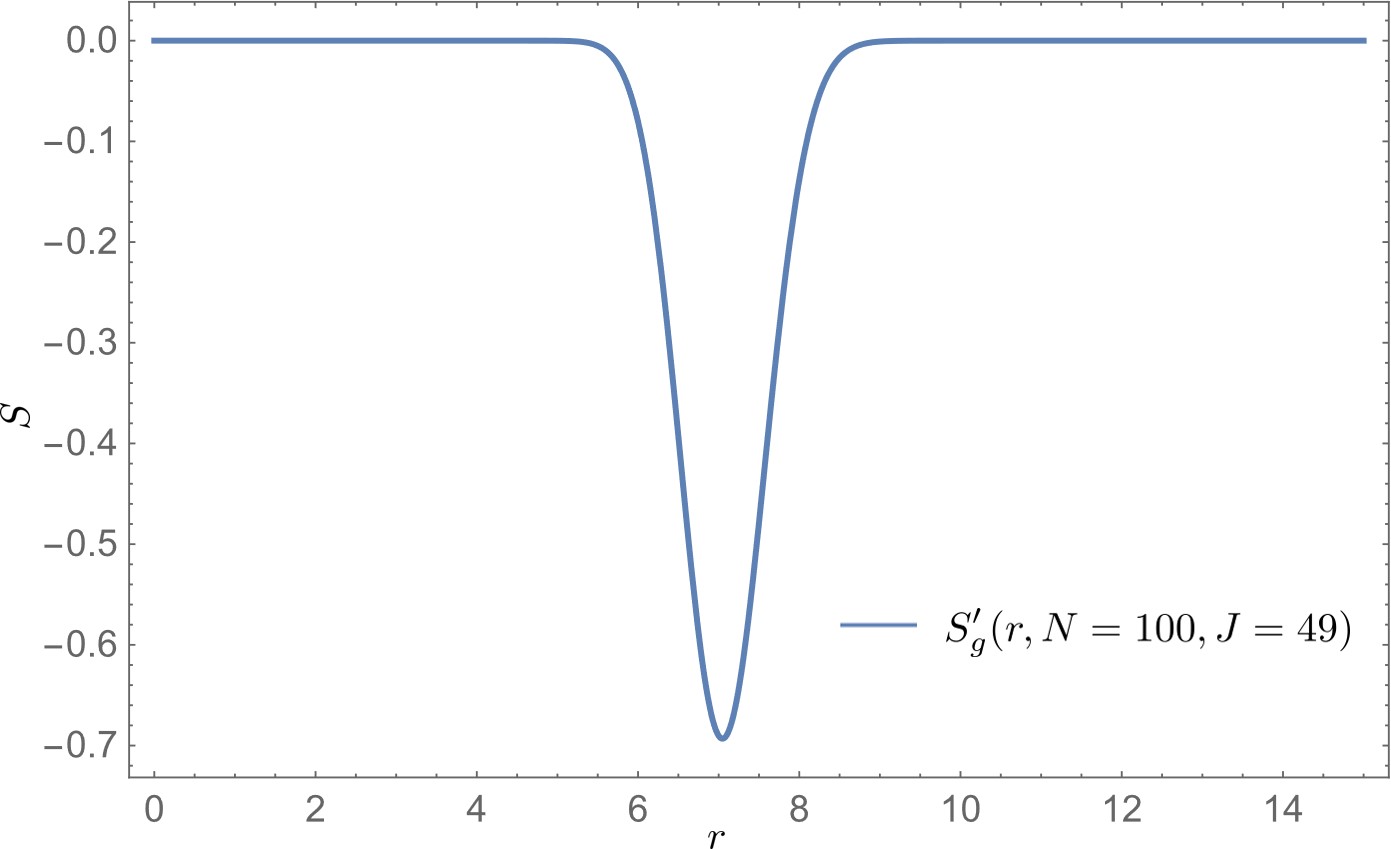}
  \caption{$S_{g}' = S_g(r,100,49)-S_0(r,100)$
  is plotted against $r$.}
  \label{fig:TSEE_d_giant}
 \end{figure}

Third, (\ref{TSEE for general Slater determinant})  for the excited state corresponding 
to a giant graviton  indicated in Figs. \ref{Droplet_giant} and \ref{level_giant} is 
\begin{align}
  S_{g}(r,N,J) = \sum^{N-2-J}_{i=0} H(a_i(r)) +\sum^{N}_{j=N-J} H(a_{j}(r)) \ .
\end{align}
In Fig. \ref{fig:TSEE_giant}, we plot $S_{g}$ for $N=100$ and $J=49$ and $S_0$ for $N=100$ against $r$.
In order to see only the contribution of the giant graviton to the target space entanglement entropy, we plot $S_{g}' = S_g(r,100,49)-S_0(r,100)$ against $r$ in Fig. \ref{fig:TSEE_d_giant}.
We see that there is a peak at $r=\sqrt{N-J}$, which is considered as the position of 
the giant graviton.


\clearpage

\section{Area of boundary in the bubbling AdS geometry}
\setcounter{equation}{0}
In this section, we calculate the area (length) of boundary of the subregion $A$ in the bubbling geometry
for the three cases, $AdS_5\times S^5$, an AdS giant graviton and a giant graviton.

We introduce the polar coordinates $(\tilde{r},\phi)$ in the $x_1-x_2$ plane.
The solution determined by a droplet in Fig. \ref{Droplet_AdS} corresponding to $AdS_5\times S^5$ is given by \cite{Lin:2004nb}
\begin{align}
  \tilde{z}(\tilde{r}, y; r_0) & \equiv z - \frac{1}{2} = \frac{\tilde{r}^2 - r_0^2 + y^2}{2 \sqrt{(\tilde{r}^2 + r_0^2 + y^2)^2 - 4 \tilde{r}^2 r_0^2} }-\frac{1}{2} \ ,\n
  V_{\phi} & = -\frac{1}{2} \left( \frac{\tilde{r}^2 + r_0^2 + y^2}{ \sqrt{(\tilde{r}^2 + r_0^2 + y^2)^2 - 4 \tilde{r}^2 r_0^2}} - 1 \right) \ , 
  \label{LLM AdS sol}
\end{align}
where $r_0$ is related to the radius of $AdS_5$ as $r_0 = R^2_{AdS}= \sqrt{N}$.

Based on (\ref{LLM AdS sol}), 
one can construct a solution for a general circular symmetric droplet as \cite{Lin:2004nb}
\begin{align}
  \tilde{z} = \sum_i (-1)^{i+1} \tilde{z}(\tilde{r}, y; r_i), \ V_{\Phi} = \sum_i (-1)^{i+1} V_{\phi} (\tilde{r}, y; r_i) \ ,
  \label{LLM gg sol.}
\end{align}
where $r_1$ is the radius of the most outer circle and $r_2$ is the radius of the second outer
circle and so on. We can construct $z$ corresponding to the droplets in Fig. \ref{Droplet_AdS_giant}
and Fig. \ref{Droplet_giant} using (\ref{LLM gg sol.}).

We identify the target space of the complex matrix model with the $x_1-x_2$ plane at $y=0$ 
and consider the subregion $A$ in the $x_1-x_2$ plane.
We calculate the area (length) of boundary of $A$ using the metric (\ref{LLM metric}).

The induced metric in the $x_1-x_2$ plane at $y=0$ denoted by $\gamma_{ij}$ is
\begin{align}
  \gamma_{\tilde{r}\tilde{r}} = h^2, \ \gamma_{\phi\phi} = - h^{-2} V_{\phi}^2 + h^2 \tilde{r}^2, \ \gamma_{\tilde{r} \phi} = \gamma_{\phi \tilde{r}} = 0  \ .
  \label{Droplet metric}
\end{align}
Then , the area (length) of the boundary of $A$ in the $x_1-x_2$ plane, which we denote by $L$, is given by
\begin{align}
  L(r) = \int_0^{2\pi} d\phi \sqrt{\gamma_{\phi\phi}} =2\pi \sqrt{ - h^{-2} V_{\phi}^2 + h^2 r^2} \ ,
  \label{circle}
\end{align}
where $h$ and $V_{\phi}$ are defined in  (\ref{func h V}).

%
First, we calculate the area of boundary of $A$ 
for $AdS_5 \times S^5$.  We see from (\ref{LLM AdS sol}) that $\gamma_{\phi\phi}$ is 
in the $y \rightarrow 0$ limit given by
\begin{align}
  \gamma_{\phi\phi} &= \lim_{y \rightarrow 0} (- h^{-2} V_{\phi}^2 + h^2 \tilde{r}^2)  
   = \frac{\tilde{r}^2 + r_0^2 - |\tilde{r}^2 - r_0^2|}{2 r_0} \ .
\end{align}
Thus, the length of the boundary of $A$ is
\begin{align}
  L(r) = \int_0^{2\pi} \sqrt{\gamma_{\phi\phi}} = \sqrt{2}\pi\sqrt{\frac{ r^2 + r_0^2 - |r^2 - r_0^2|}{r_0}} \ .
  \label{Length}
\end{align}
$L$ is proportional to $r$ for $r<r_0$: $L(r) = \sqrt{\frac{2}{r_0}}\pi r$.
This behavior qualitatively agrees with that of the target space entanglement entropy in
Fig. \ref{fig:TSEE_ground_state}.
$L$ is constant for $r>r_0$,: $L(r) = \sqrt{2r_0}\pi $.


Next, we calculate the area of the  boundary of $A$ for the (AdS) giant gravitons.
We see from (\ref{LLM AdS sol}) and (\ref{LLM gg sol.}) that 
\begin{align}
  z &= \tilde{z} + \frac{1}{2} = \sum_{i=1}^{3} (-1)^{i+1} \tilde{z}(r, y; r_i) + \frac{1}{2}  \ , \n
  V_{\phi}& = \sum_{i=1}^{3} (-1)^{i+1} V_{\phi} (r, y; r_i)  \ ,
\end{align}
where $r_1 > r_2 > r_3$.
By using (\ref{Droplet metric}), we can calculate $\gamma_{\phi\phi}$ in the $y \rightarrow 0$ limit. 
The results are summarized in appendix \ref{sec:appendix B}.
We denote the area of boundary of $A$ for $AdS_5\times S^5 $, an AdS giant graviton and a 
giant graviton by $L_0(r)$. $L_{AdS}(r)$ and $L_g(r)$, respectively.

\begin{figure}[t]
  \centering
  \includegraphics[width=11cm]{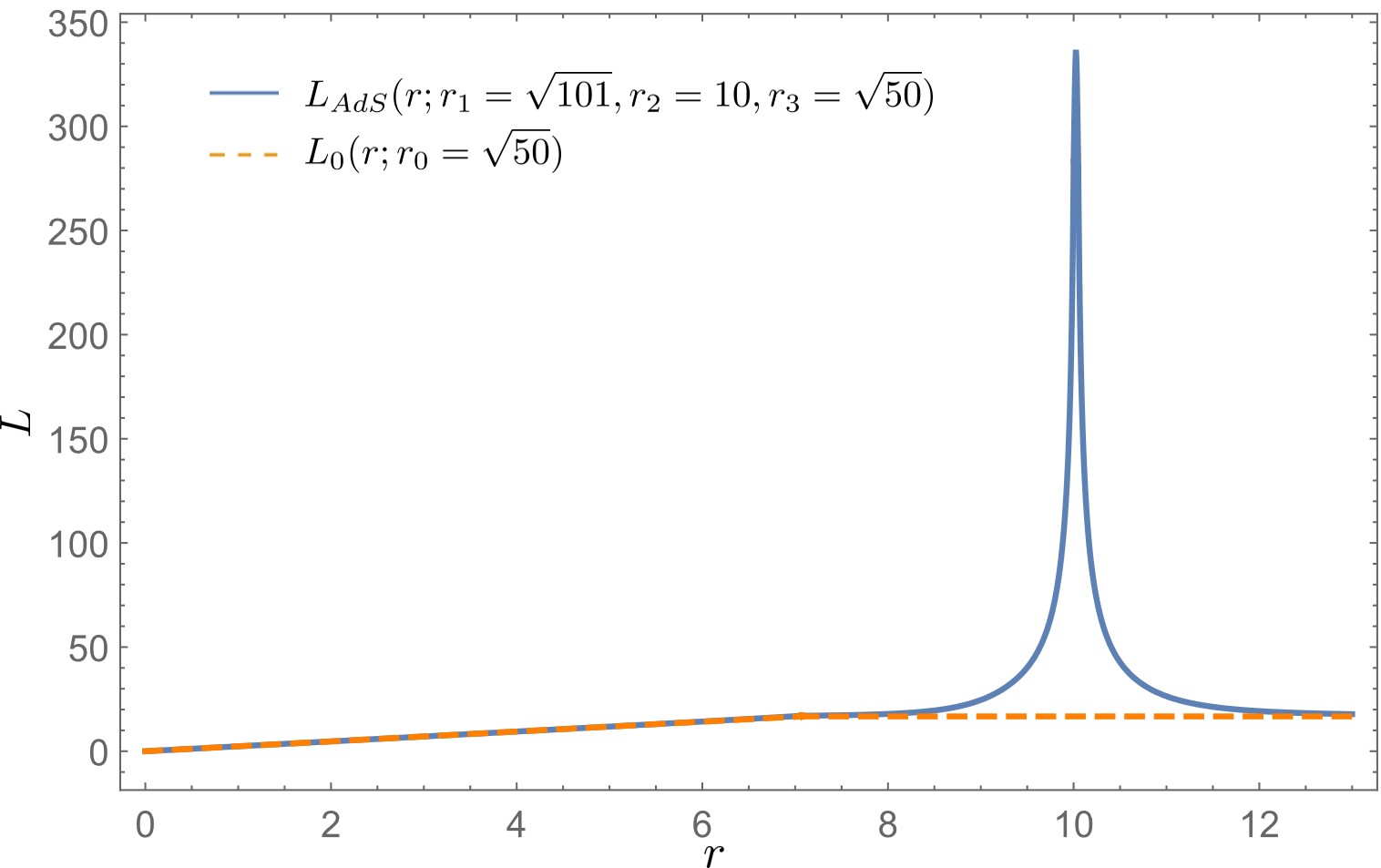}
  \caption{The solid line represents the area of the boundary of the subregion $A$ for an AdS giant
 graviton $L_{AdS}$ with $r_1 = \sqrt{101}, r_2 = 10$ and $ r_3=\sqrt{50}$, while the dashed line represents
 that for the ground state
 $L_{0}$ with $r_0= \sqrt{50}$.}
  \label{fig:Area_ads_giant}
 \end{figure}

\begin{figure}[t]
  \centering
  \includegraphics[width=11cm]{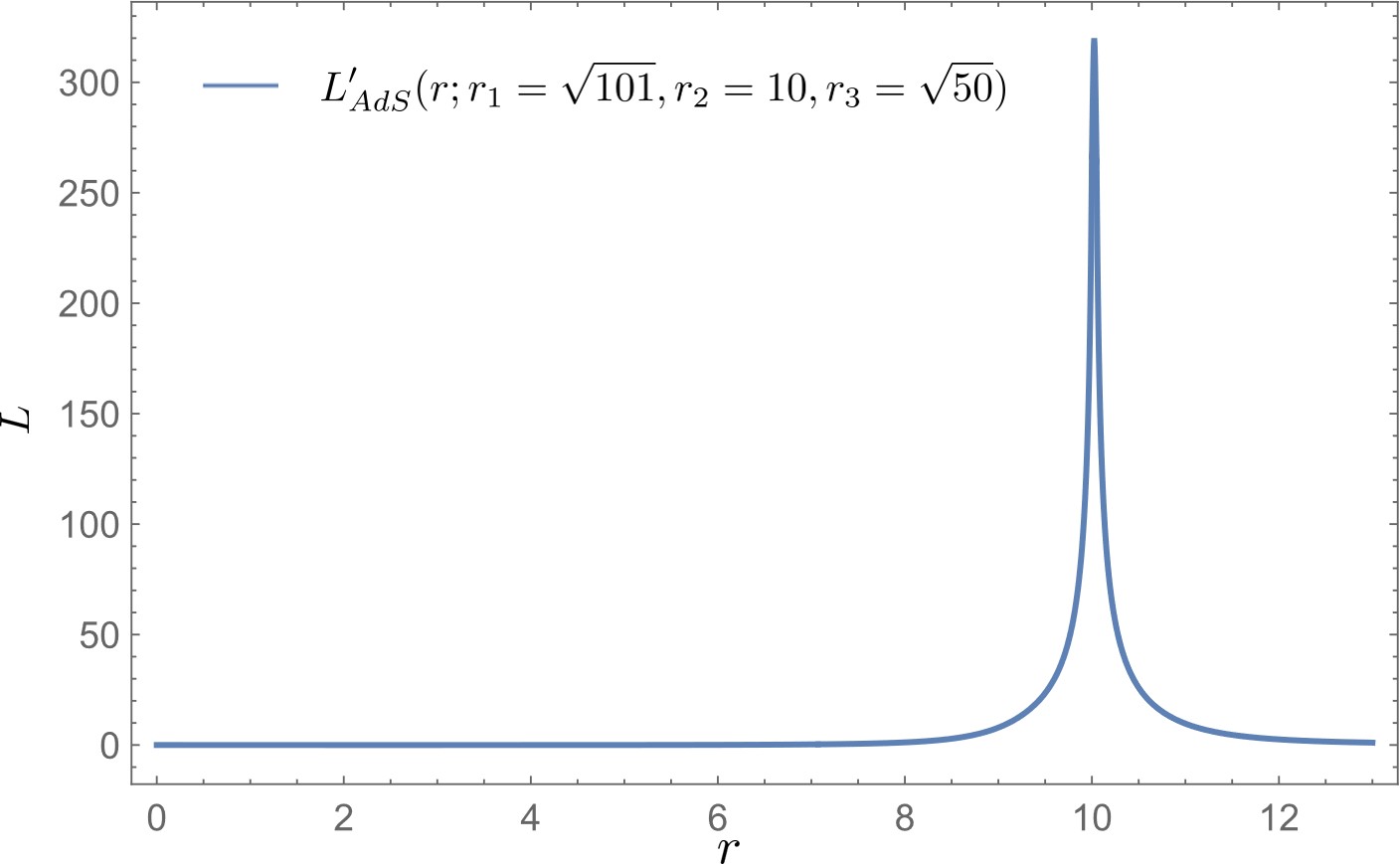}
  \caption{$L_{AdS}' = L_{AdS}(r)-L_0(r)$ is plotted against $r$.}
  \label{fig:Area_d_ads_giant}
 \end{figure}

In Fig. \ref{fig:Area_ads_giant}, we plot $L_{AdS}(r)$ for $r_1 = \sqrt{101}, r_2 = 10, r_3=\sqrt{50}$
and $L_0(r)$ for $r_0 = r_3$ against $r$. 
As in the target space entanglement entropy, we subtract $L_0$ from $L_{AdS}$.
In Fig .\ref{fig:Area_d_ads_giant}, we plot $L_{AdS}' = L_{AdS}(r)-L_0(r)$ against $r$.
We see that $L_{AdS}'$ has a peak around $r = r_1 = \sqrt{N}$.
This behavior of $L_{AdS}'$ qualitatively agrees with $S_{AdS}'$ in Fig. \ref{fig:TSEE_d_ads_giant}.

\begin{figure}[t]
  \centering
  \includegraphics[width=11cm]{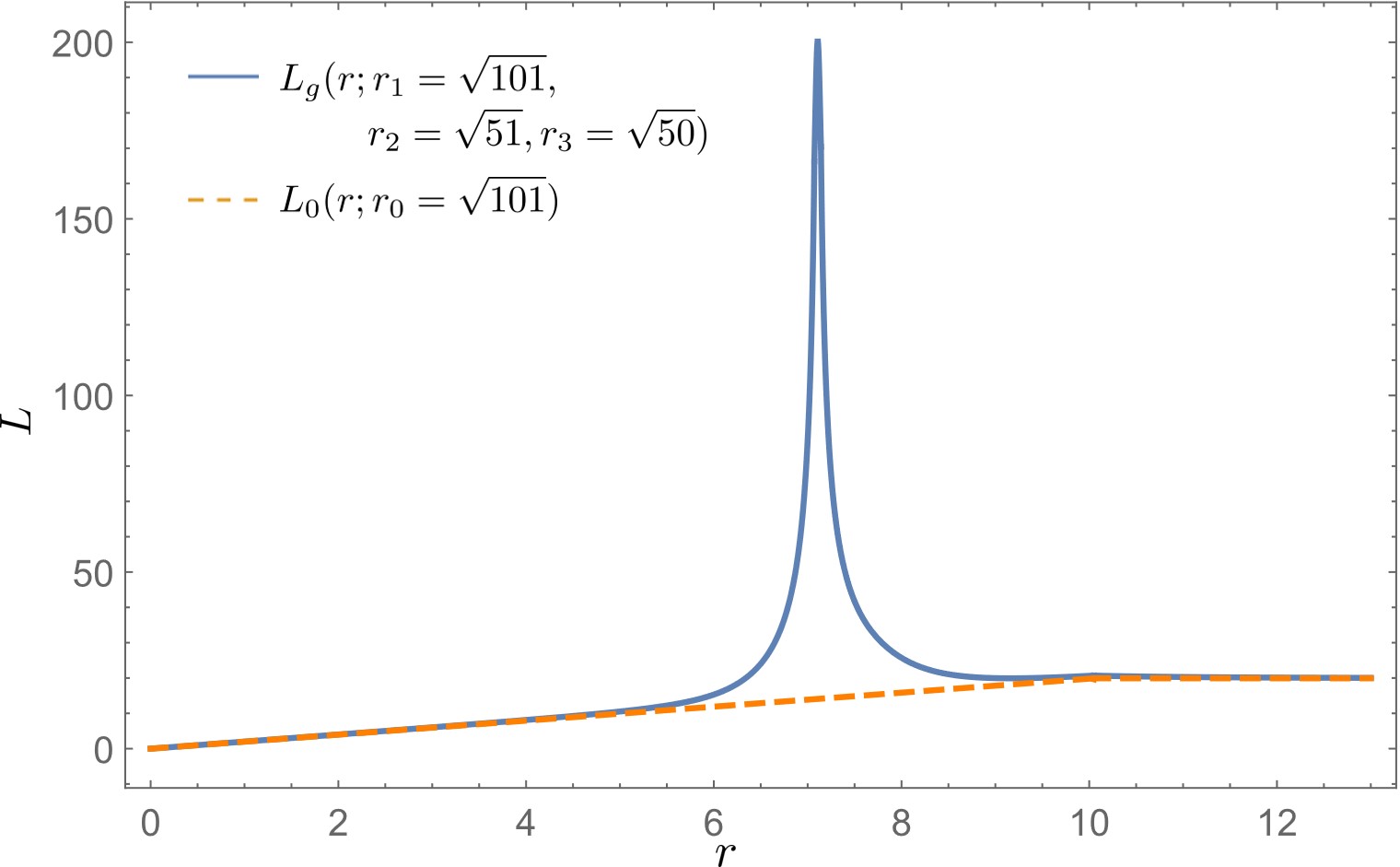}
  \caption{The solid line represents the area of boundary of subregion $A$ for a
  giant graviton $L_{AdS}$ with $r_1 = \sqrt{101}, r_2 = \sqrt{51}$ and $ r_3=\sqrt{50}$,
  while tha dashed line represents that for the ground state $L_{0}$ with 
  $r_0= \sqrt{101}$.}
  \label{fig:Area_giant}
 \end{figure}
\begin{figure}[t]
  \centering
  \includegraphics[width=11cm]{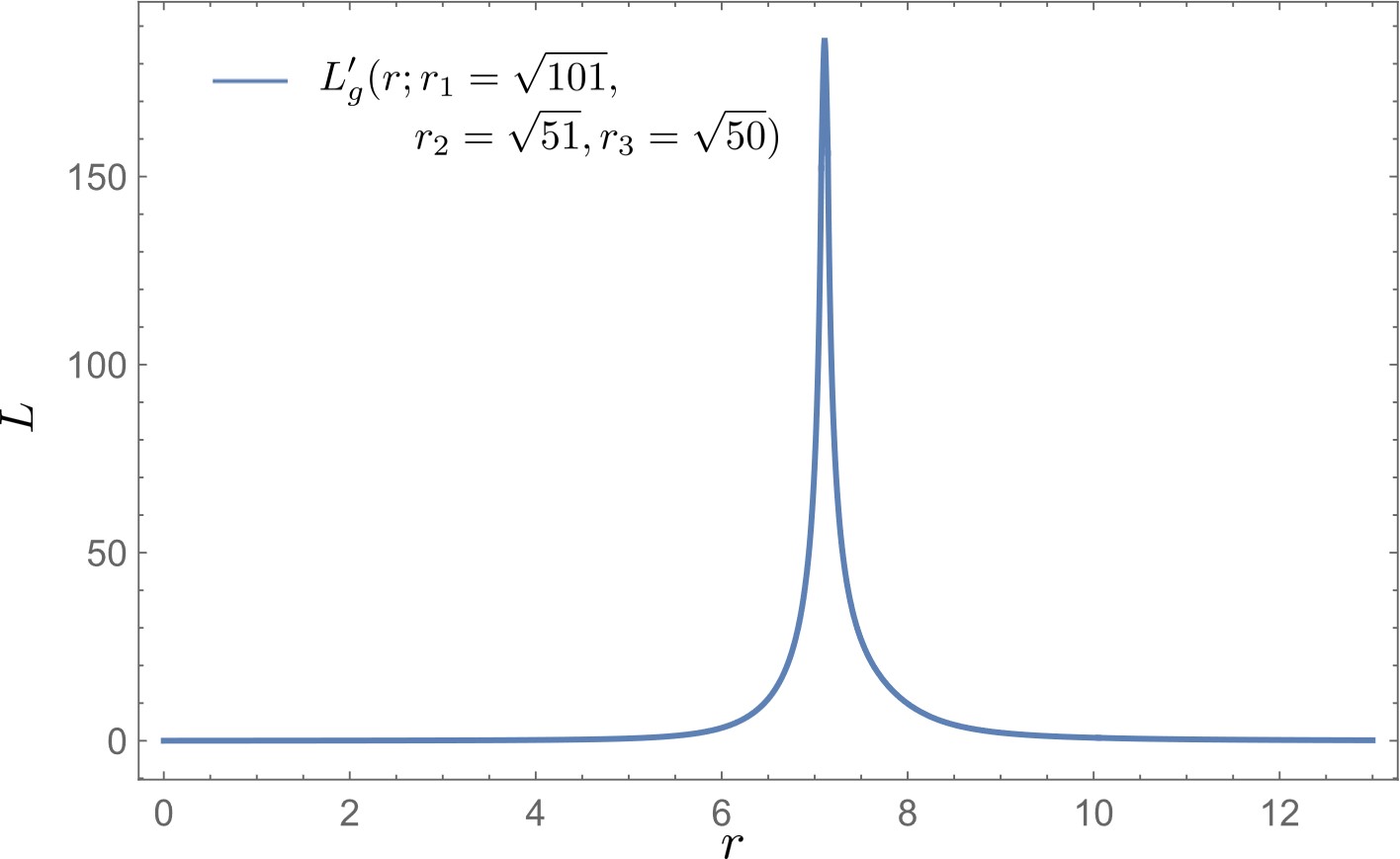}
  \caption{$L_{g}' = L_{g}(r)-L_0(r)$ is plotted against $r$.}
  \label{fig:Area_d_giant}
 \end{figure}

In Fig. \ref{fig:Area_giant}, we plot $L_{g}(r)$ for $r_1 = \sqrt{101}, r_2 = \sqrt{51}, r_3=\sqrt{50}$
and $L_0$ for $r_0 = r_1$  against $r$. 
In Fig. \ref{fig:Area_d_giant}, we plot $L_{g}' = L_{g}(r)-L_0(r)$ against $r$.
We see that $L_{g}'$ has a peak around $r = r_2 = \sqrt{N}$.
The sign of the peak is different from that of the peak  in  Fig. \ref{fig:TSEE_d_giant}.:
the area of boundary of $A$ has a positive peak, while the target space 
entanglement entropy a negative peak.
The peak of the target space entanglement for the giant graviton comes from disappearance of
a fermion. If we reinterpret this as a particle corresponding to a  hole, the sign of the peak should flip.

In this way, we find that
the contribution of the (AdS) giant gravitons to the area (length) of  boundary of $A$ qualitatively agrees with
that to the target space entanglement entropy.


 
 \clearpage

\section{Conclusion and discussion}
\setcounter{equation}{0}
In this paper, we studied the target space entanglement entropy in 
the complex matrix model that describes the chiral primary sector in $\mathcal{N}=4$ SYM on
$R \times S^3$. The eigenvalues of the complex matrix model are identified with
the position coordinates of  fermions in two-dimensional plane, and the eigenvalue distribution
gives a droplet formed by the fermions, which is identified with that in the bubbling geometry
specifying a boundary condition for a half-BPS solution with $R \times SO(4) \times SO(4)$ in type IIB
supergravity. We calculated the target space entanglement entropy of a subregion in two-dimensional
plane with droplets corresponding to $AdS_5 \times S^5$, an AdS giant graviton and a giant graviton
in the bubbling geometry. We also calculated the area of boundary of the subregion in the bubbling 
geometry, and found a qualitative agreement between the target space entanglement entropy
and the area of boundary.
For the following reason, it is considered as reasonable that 
we subtracted the contribution of the ground state ($AdS_5\times S^5$) when we compared
the target space entanglement entropy and the area of boundary for the (AdS) giant gravitons.
The fermions forming the droplet corresponding to the ground state
originate from D3-branes in flat space-time.
These D3-branes disappear, and their presence is trade for a deformation of flat space-time to $AdS_5\times S^5$. The target space entanglement entropy for the ground state
measures the entanglement among the fermions corresponding to the disappearing D3-branes.

In order to see whether a quantitative agreement exists, we need to identify the effective Newton constant 
in (2+1)-dimensional space-time consisting the two-dimensional plane and the time and
fix the factor of proportionality $1/4G_N$ between the target space entanglement entropy and the area.
Possibly, we also need more elaborated correspondence between chiral primary states and droplets.
Finally, we make comments on the scaling with respect to $N$. 
The maximum of the target space entanglement entropy 
of a subregion in two-dimensional plane with a droplet corresponding
to $AdS_5\times S^5$ scales almost as $N^{1/2}$
as seen in Fig. \ref{fig:TSEE_ground_state}, while
the maximum of the magnitude of the subtracted one corresponding to an AdS giant graviton or a giant graviton scales as $N^0$.
On the other hand, we see from (\ref{Length}) that
the maximum of the area of boundary for
$AdS_5\times S^5$ scales as $\sqrt{r_0} \sim N^{1/4}$,
and we have verified that the maximum
of the subtracted one for an AdS giant graviton or a giant graviton behaves
almost as $N^{3/4}$ when the ratio between $r_2$ and $r_3$ for an AdS giant graviton in Fig. \ref{Droplet_AdS_giant}
and $r_1$ and $r_2$ for a giant graviton in Fig. \ref{Droplet_giant} is fixed. 
This difference of the relative ratios between the maxima
for $AdS_5\times S^5$ and (AdS) giant gravitons does not matter
if only subtracted quantities make physical sense as discussed above.
The fact that the entanglement entropy and the area  scales as $N^0$ and
$N^{3/4}$, respectively, commonly for an AdS giant gravion and a giant graviton
suggests that the difference between the scalings of
the subtracted
entropy and area can be compensated by 
the scaling of the effective Newton constant.
While it is conjectured in \cite{Das:2020jhy,Das:2020xoa} 
that the target space entanglement entropy
in 9+1 dimensions scales as $N^2$, the target space entanglement entropy
calculated in this paper is defined in a (2+1)-dimensional subspace
so that it possibly scales in a different manner. 
We hope to report progress in these issues in  the near future.

\section*{Acknowledgments}
A.T. was supported in part by Grant-in-Aid for Scientific Research (No. 18K03614 and No. 21K03532) from
Japan Society for the Promotion of Science.
K.Y. was supported in part by Grant-in-Aid for JSPS
Fellows (No. 20J13836).


\appendix

\renewcommand{\theequation}{A.\arabic{equation}}

\section{Harmonic oscillator in two-dimensional plane}
\renewcommand{\theequation}{A.\arabic{equation}}
\setcounter{equation}{0}
In this appendix, we review that the wave function of the lowest Landau level arises as a wave function
of holomorphic sector of a particle under harmonic oscillator potential in two-dimensional plane..
The hamiltonian  is represented in the complex coordinate $(z,z^*)$ basis as
\begin{align}
  \hat{h} = - \frac{\del}{\del z}\frac{\del}{\del z^*} + zz^* = \hat{c}^{\dagger}_1 \hat{c}_1 + \hat{c}^{\dagger}_2 \hat{c}_2 +1   \ ,
\end{align}
where
\begin{align}
  &\hat{c}_1 \equiv \frac{1}{\sqrt{2}} \left(z + \frac{\del}{ \del z^*} \right),  \; \; \hat{c}_2 = \frac{1}{\sqrt{2}} \left(z^* + \frac{\del}{ \del z} \right) \n
  &\hat{c}^{\dagger}_1 \equiv \frac{1}{\sqrt{2}} \left(z^* - \frac{\del}{ \del z} \right), \; \;  \hat{c}^{\dagger}_2 = \frac{1}{\sqrt{2}} \left(z - \frac{\del}{ \del z^*}\right)
\end{align}
The commutation relations of $\hat{c}_i$ and $\hat{c}_i^{\dagger}$ are given by
\begin{align}
  [\hat{c}_i,  \hat{c}^{\dagger}_j]=\delta_{ij}, \;\; [\hat{c}_i, \hat{c}_j]= [ \hat{c}^{\dagger}_i,  \;\;
 \hat{c}^{\dagger}_j]=0
\end{align}
The wave function of the ground state,  $\Phi_{0,0}(z, z^*)$,
satisfies $\hat{c} \Phi_{0,0} = 0$, which gives
\begin{align}
  \Phi_{0,0} = \frac{1}{\sqrt{\pi}} e^{-z z^*}   \ .
\end{align}
The wave functions of excited states are given by
\begin{align}
  \Phi_{k, l} (z, z^*) = \frac{1}{\sqrt{k! l!}} (\hat{c}^{\dagger}_1)^k (\hat{c}^{\dagger}_2)^l \Phi_{0,0}  \ .
\end{align}
The energy eigenvalue of the above states are $k+l+1$.
In particular, the wave functions of the $(0,l)$ excited states, 
$\Phi_{l}(z, z^*) \equiv \Phi_{0,l}(z, z^*)$, take a holomorphic form
\begin{align}
  \Phi_{l}(z, z^*)= \sqrt{\frac{2^l}{l! \pi}} z^l e^{-z z^*}  \ .
\end{align}
$\Phi_{l}(z, z^*)$ are the wave functions of the lowest Landau level in the 
context of quantum Hall effect.

\section{$\gamma_{\phi \phi}$ for (AdS) giant gravitons}
\label{sec:appendix B}
\renewcommand{\theequation}{B.\arabic{equation}}
\setcounter{equation}{0}

$\gamma_{\phi\phi}$ is given in the $y \rightarrow 0$ limit as follows:

\noindent
(i) $r>r_1$
\begin{align}
  &\lim_{y \rightarrow 0} \gamma_{\phi\phi}
  =\frac{r_1^2(r^6 + r_2^2 (r^4 + 3 r_3 \tilde{r}^2) - r_3^2 (4 r^4 + r_2^4)) - r^4(\tilde{r}^2 + r_2^2)(r_2^2 - r_3^2)}{(\tilde{r}^2 - r_1^2) (\tilde{r}^2 - r_2^2)^2 (\tilde{r}^2 - r_3^2) \sqrt{\frac{r_1^2}{(\tilde{r}^2 - r_1^2)^2} - \frac{r_2^2}{(\tilde{r}^2 - r_2^2)^2} + \frac{r_3^2}{(\tilde{r}^2 - r_3^2)^2}}}  \ .
\end{align}

\noindent
(ii) $r_1>r>r_2$
\begin{align}
  &\lim_{y \rightarrow 0} \gamma_{\phi\phi}
  = \frac{-\tilde{r}^2 (\tilde{r}^2+ r_3^2)(\tilde{r}^2(\tilde{r}^2 - 4 r_3^2) + r_1^2 (r_2^2 - r_3^2) - r_2^2 r_3^2)}{(\tilde{r}^2 - r_1^2) (\tilde{r}^2 - r_2^2) (\tilde{r}^2 - r_3^2)^2 \sqrt{{\frac{r_1^2}{(\tilde{r}^2 - r_1^2)^2} + \frac{r_2^2}{(\tilde{r}^2 - r_2^2)^2} - \frac{r_3^2}{(\tilde{r}^2 - r_3^2)^2}}}}  \ .
\end{align}

\noindent
(iii) $r_2>r>r_3$
\begin{align}
  &\lim_{y \rightarrow 0} \gamma_{\phi\phi}
  = \frac{\tilde{r}^2 r_1^2(r^4 - \tilde{r}^2(r_2^2 + r_3^2)- 3r_2^2 r_3^2 ) - r^6 (r_2^2 + r_3^2) +4 r^4 r_2^2 r_3^2 - r_1^4(r_2^2 r_3^2 + r^4) }{(\tilde{r}^2 - r_1^2)^2 (\tilde{r}^2 - r_2^2) (\tilde{r}^2 - r_3^2) \sqrt{-\frac{r_1^2}{(\tilde{r}^2 - r_1^2)^2} + \frac{r_2^2}{(\tilde{r}^2 - r_2^2)^2} + \frac{r_3^2}{(\tilde{r}^2 - r_3^2)^2}}} \ .
\end{align}

\noindent
(iv) $r_3>r$
\begin{align}
  &\lim_{y \rightarrow 0} \gamma_{\phi\phi}
  = \frac{\tilde{r}^2(- r^6 + \tilde{r}^2 r_2^2(3 \tilde{r}^2 + r_3^2) +(r_2^2 - r_3^2)(\tilde{r}^2 + r_2^2) - r_2^4 (4\tilde{r}^2 - r_3^2) )}{(\tilde{r}^2 - r_1^2) (\tilde{r}^2 - r_2^2)^2 (\tilde{r}^2 - r_3^2) \sqrt{\frac{r_1^2}{(\tilde{r}^2 - r_1^2)^2} - \frac{r_2^2 }{(\tilde{r}^2 - r_2^2)^2} + \frac{r_3^2}{(\tilde{r}^2 - r_3^2)^2}}}  \ .
\end{align}



\end{document}